\newcommand{\beq}{\begin{eqnarray}}
\newcommand{\eeq}{\end{eqnarray}}

\newcommand{\eps}{\epsilon}
\newcommand{\half}{{1\over 2}}
\newcommand{\e}{{\rm e}}

\renewcommand{\theequation}{\thesection.\arabic{equation}}

\newcommand{\cl}{\centerline}

\newcommand{\btem}{\bibitem}

\newcommand{\TK}{T.\ Kunihiro}
\newcommand{\PL}{Phys.\ Lett.\ {\bf B}}
\newcommand{\PTP}{Prog.\ Theor.\ Phys.}

\newcommand{\PRL}{Phys.\ Rev. \ Lett.}

\documentstyle[12pt]{article}

\begin{document}
\begin{flushright}
RYUTHP-95/3 \\
May, 1995 \\
\end{flushright}

\begin{center}
{\large {\bf  A Geometrical Formulation of the Renormalization Group
Method for Global  Analysis}} \\

\end{center}

\vspace{1cm}

{\cl {Teiji Kunihiro}}

\bigskip

\begin{center}

Faculty of Science and Technology, Ryukoku University,\\
Seta, Ohtsu-city, 520-21, Japan\\
\end{center}

\begin{abstract}
On the basis of the classical theory of envelope,
 we formulate
 the renormalization  group (RG) method  for global analysis,
 recently proposed by
 Goldenfeld et al.
 It is clarified why the RG equation improves things.
  \end{abstract}

\section{Introduction}
\renewcommand{\theequation}{\thesection.\arabic{equation}}
Recently, Goldenfeld et al\cite{goldenfeld} have proposed a new method
based on the
 renormalization group (RG) equation \cite{rg,JZ} to get the asymptotic
 behavior of solutions of differential equations.
The method is simple  and has a wide
 variety of applications including singular and reductive perturbation
 problems in a unified way.  However, the reason is
 obscure why the RG equation can be relevant
 and useful for global analysis: The RG equation is usually
 related with the scale invariance of the system under consideration.
The equations which can be treated in the RG method are not confined to
those with
 scale invariance\cite{goldenfeld}.
 Actually, what the RG method does in \cite{goldenfeld}
 may be said to
 construct an approximate but  {\em global} solution from
 the ones with a local nature which were obtained in the perturbation
theory; the RG equation is used to improve the global behavior of the
 local solutions.
This fact suggests that  the RG method   can be  formulated
 in a purely mathematical way  without recourse to
 the concept of the RG.
A purpose of the paper is to show that this is the case,
 thereby   reveal the
 mathematical structure of the method.

Our formulation is based on the
 classical theory of envelopes \cite{courant}.
 As everybody knows, the envelope
of a family of curves or surfaces  has usually an improved  global nature
compared with  the curves or surfaces in the family.
 So it is natural that the theory of envelopes may have  some power
 for global analysis.
One will recognize that  the powerfulness of the
 RG equation in  global analysis and also in
 the quantum field theory\cite{rg, JZ}
is due to the fact that it is essentially  an envelope equation.
We shall also give a proof
 as to why the RG equation can give a globally improved
 solution to  differential equations.

In the next section, a short review is given on the classical
theory of envelopes,
 the notion of which is essential for the understanding of the present
paper.  In \S 3,we formulate the RG method in the
 context of the theory of envelopes and give a foundation to the method.
 In \S 4,  we show a couple of other examples to apply our formulation.
The last section is devoted to a brief summary and concluding remarks.

\section{A short review of the classical theory of  envelopes}
\renewcommand{\theequation}{\thesection.\arabic{equation}}

To make the discussions  in the following sections clear,
 we here give a brief review of the theory of  envelopes.
Although the theory can be formulated in higher
dimensions\cite{courant},
 we take here only the one-dimensional envelopes. i.e., curves,
 for simplicity.

Let $\{C_{\tau}\}_{\tau}$ be a family of curves  parametrized by $\tau$
in the $x$-$y$  plane; here $C_{\tau}$ is  represented by the
 equation
\beq
F(x, y, \tau)=0.
\eeq
 We suppose
 that $\{C_\tau\}_{\tau}$ has the envelope $E$, which is represented by the
 equation
\beq
G(x, y)=0.
\eeq
The problem is to obtain $G(x, y)$ from $F(x, y,\tau)$.

Now let $E$ and a curve $C_{\tau_0}$ have the common tangent line at
 $(x,y)=(x_0,y_0)$, i.e., $(x_0, y_0)$ is the point of tangency.
  Then $x_0$ and $y_0$ are functions of $\tau_0$;
$x_0=\phi(\tau_0),\ y_0= \psi(\tau_0)$, and of course $ G(x_0, y_0)=0$.
Conversely, for each point $(x_0, y_0)$ on $E$, there exists a  parameter
 $\tau_0$.
 So we can reduce the problem to get $\tau_0$ as a function of
$(x_0,y_0)$;
 then $G(x, y)$ is obtained as
$F(x, y, \tau(x,y))=G(x,y)$.\footnote{Since there
 is a relation $G(x_0, y_0)=0$ between $x_0$ and $y_0$, $\tau_0$ is actually a
 function of $x_0$ {\em or } $y_0$.} $\tau_0(x_0, y_0)$ can be obtained as
 follows.

 The tangent line of $E$ at $(x_0, y_0)$  is given by
\beq
\psi '(\tau_0)(x-x_0)- \phi'(\tau_0)(y-y_0)=0,
\eeq
 while the tangent line of $C_{\tau_0}$ at the same point reads
\beq
F_x(x_0, y_0, \tau_0)(x-x_0)+ F_y(x_0, y_0, \tau_0)(y-y_0)=0.
\eeq
Here $F_x=\partial F/\partial x$ and $F_y=\partial F/\partial y$.
Since both equations must give the same line,
\beq
F_x(x_0, y_0, \tau_0)\phi'(\tau_0)+ F_y(x_0, y_0, \tau_0)\psi'(\tau_0)=0.
\eeq
 On the other hand, differentiating $F(x(\tau_0), y(\tau_0), \tau_0)=0$ with
respect
to $\tau_0$, one has
\beq
F_x(x_0, y_0, \tau_0)\phi'(\tau_0)+ F_y(x_0, y_0, \tau_0)\psi'(\tau_0)
 + F_{\tau_0}(x_0, y_0, \tau_0)=0,
\eeq
hence
\beq
 F_{\tau_0}(x_0, y_0, \tau_0)
\equiv\frac{\partial F(x_0, y_0, \tau_0)}{\partial \tau_0}=0.
\eeq

One can thus
 eliminate the parameter $\tau_0$ to get a relation between  $x_0$ and $y_0$,
\beq
G(x, y) = F(x, y, \tau_0(x, y))=0,
\eeq
 with the replacement $(x_0, y_0)\rightarrow (x,y)$. $G(x, y)$ is called
 the discriminant of $F(x, y, t)$.

Comments are in order here:
(i)\  When the family of curves is given by the
 function $ y=f(x, \tau)$, the condition Eq.(2.7) is reduced to
${\partial f}/{\partial \tau_0}=0$; the envelope is given by
 $y=f(x, \tau_0(x))$.
(ii)\  The equation $G(x, y)=0$ may give
 not only the envelope $E$ but also a set of
singularities of the curves $\{C_{\tau}\}_{\tau}$. This is because
 the condition that
$\partial F/\partial x = \partial F/\partial y=0$ is also
 compatible with  Eq. (2.7).

As an example, let
\beq
y=f(x, \tau)= {\rm e}^{-\eps \tau}(1 - \eps\cdot(x -\tau)) + {\rm e}^{-x}.
\eeq
Note that $y$ is unbound for $x -\tau \rightarrow \infty$ due to the
 secular term.

The envelope $E$ of the curves $C_{\tau}$ is obtained as follows:
 From ${\partial f}/{\partial \tau}=0$, one has $\tau=x$.
That is, the parameter in this case is the $x$-coordinate of the
 point of the tangency of  $E$ and  $C_\tau$.
Thus the envelope is found to be
\beq
y=f(x, x)=  {\rm e}^{-\eps x} - {\rm e}^{-x}.
\eeq
One can see that the envelope is bound even for $x\rightarrow \infty$.
 In short, we have obtained a function as the envelope with a
 better global nature from functions which are bound only locally.

As an illustration, we show in Fig.1 some of
the curves given by $y=f(x, \tau_0)$ together with  the envelope.

\vspace{1cm}
{\cl {\fbox{\bf Fig.1}}}
\setcounter{equation}{0}
\section{Formulation of the RG method based on the theory of envelopes}
\renewcommand{\theequation}{\thesection.\arabic{equation}}

In this section, we formulate and give a foundation of
the RG method\cite{goldenfeld} in the
 context of the classical theory of envelopes sketched in the previous section.
Our formulation also includes an improvement of the prescription.


Although the RG method  can be
 applied to both (non-linear) ordinary and partial differential
equations, let us take the following simplest example to show
our formulation:
\beq
\frac{d^2 x}{dt^2}\ +\ \eps \frac{dx}{dt}\ +\ x\ =\ 0,
\eeq
where $\eps$ is supposed to be small. The solution to Eq.(3.1) reads
\beq
 x(t)= A \exp (-\frac{\eps}{2} t)\sin( \sqrt{1-\frac{\eps^2}{4}} t + \theta),
\eeq
where $A$ and $\theta$ are constant to be determined by an initial
 condition.

 Now, let us blindly try to get the solution in the  perturbation theory,
  expanding $x$ as
\beq
x(t) = x_0(t) \ +\ \eps  x_1(t)\ +\ \eps ^2 x_2(t)\ +\ ... ,
\eeq
 where $x_n$ ($n= 0, 1, 2 ...$) satisfy
\beq
\ddot{x}_0  +\ x_0\ =\ 0,
\ \ \ \
 \ddot{x}_{n+1}\ +\ x_{n+1}\ =- \dot{x}_n .
\eeq
Thus $x_0 = A_0\sin (t +\theta_0)$, and then
$\ddot{x}_1\ +\ x_1\ =- A_0\cos(t+\theta_0),$
 and so on. Then we get  for $x_1$ and $x_2$ as special solutions
\beq
x_1(t)&=& -\frac{A_0}{2}\cdot (t -t_0)\sin(t+\theta_0),\nonumber \\
 x_2(t)& =&
 \frac{A_0}{8}\{ (t-t_0)^2\sin(t +\theta_0) - (t-t_0)\cos(t+\theta_0)\}.
\eeq
Here we have intentionally omitted the unperturbed solution from
  $x_n(t)\ (n= 1, 2,...)$. Although this prescription is not adopted in
\cite{goldenfeld}, the proceeding calculations are
 simplified with this prescription; see
 also \S 4.\footnote{It is amusing to see that the unperturbed solution
 in the higher order terms $x_n$\ ($n=1, 2, ...$) is analogous to
 the ``dangerous'' term in the Bogoliubov's sense
in the quantum-filed theory of
 superfluidity and superconductivity\cite{danger}.}
It should be noted  that  the secular terms have appeared
 in the higher order terms, which are  absent in the
exact solution and invalidates the perturbation theory for $t$ far
 away from $t_0$.

Inserting Eq.(3.5) into Eq.(3.3), we have
\beq
x(t, t_0)&=& A_0\sin (t +\theta_0) -\eps\frac{A_0}{2} (t -t_0)\sin(t+\theta_0)
  \nonumber \\
 \ \ \ & \ \ \ & +\eps^2\frac{A_0}{8}
\{ (t-t_0)^2\sin(t +\theta_0) - (t-t_0)\cos(t+\theta_0)\}
  + O(\eps^3)
\eeq

Now we have a family of curves $\{C_{t_0}\}_{t_0}$ given by functions
$\{x(t, t_0)\}_{t_0}$ parametrized with $t_0$.
 They are all
 solutions of Eq. (3.1) up to $O(\eps ^3)$, but
 only valid locally, i.e., for $t$ near $t_0$.
 Let us  find  a function $x_{_E}(t)$ representing the envelope $E$ of
$\{C_{t_0}\}_{t_0}$.

According to the previous section,
 we only have to eliminate $t_0$ from
\beq
 \frac{\partial x(t, t_0)}{\partial t_0}=0,
\eeq
 and insert the resultant $t_0(t)$ into $x(t, t_0)$.
Then we identify as $x_{_E}(t)=x(t, t_0(t))$.
It will be shown  that  $x_{_E}(t)$ satisfies the original differential
equation  Eq. (3.1) {\em uniformly} for $\forall t$ up to
 $O(\eps^4)$; see below.

Eq.(3.7) is in the same form as  the RG equation, hence the name of
 the RG method\cite{goldenfeld}.
In our formulation, this is a condition for constructing
 the envelope.

Here comes another crucial point of the method.
We assume  that $A_0$ and $B_0$ are functionally dependent on $t_0$;
\beq
A_0=A_0(t_0), \ \ \ \ \theta_0=\theta_0(t_0),
\eeq
accordingly $x(t, t_0)= x(t, A_0(t_0), \theta_0(t_0), t_0)$.  Then
it will be  found that Eq. (3.7) gives a complicated equation
involving  $A_0(t_0), \theta_0(t_0)$ and  their derivatives as well as $t_0$.
It turns out, however, that one can  actually
 greatly reduce the complexity of  the equation
by assuming that the parameter $t_0$ coincides with the point of tangency,
 that is ,
\beq
t_0 =t\ ,
\eeq
because  $A_0(t_0)$ and $\theta_0(t_0)$ can be determined so
that $t_0 =t$.
 We remark here that  the meaning of setting $t_0=t$ is not
clearly explained in \cite{goldenfeld}, while in our case, the setting has
 the  clear meaning to choose the point of tangency at
$t=t_0$.\footnote{It is interesting that the procedure to get the envelope
 of $x(t, A_0(t_0), \theta_0(t_0), t_0)$ assuming a functional dependence of
$A_0$ and $\theta_0$ on $t_0$ is similar to the standard prescription
  in which the general
solution of a partial differential equation of first order is constructed
 from the complete solution.\cite{courant}}

{}From Eq.'s (3.7)  and (3.9), we have
\beq
\frac{dA_0}{dt_0} + \eps A_0 =0,  \ \ \
\frac{d\theta_0}{dt_0}+\frac{\eps^2}{8}=0.
\eeq
Solving the simple equations, we have
\beq
A_0(t_0)= \bar{A}{\rm e}^{-\eps t_0/2}, \ \ \
\theta _0(t_0)= -\frac{\eps^2}{8}t_0 + \bar{\theta},
\eeq
where $\bar{A}$ and $\bar{\theta}$ are constant numbers.
 Thus we get
\beq
x_{_E}(t)=x(t,t)=
\bar{A}\exp(-\frac{\eps}{2} t)\sin((1-\frac{\eps ^2}{8})t +
\bar{\theta}).
\eeq
Noting that $\sqrt{1 - {\eps^2}/{4}}= 1 - {\eps^2}/{8} + O(\eps ^4)$,
 one finds
 that the resultant envelope function $x_{_E}(t)$ is an approximate but
{\em  global} solution to Eq.(3.1); see Eq. (3.2).
In short, the solution obtained in the perturbation theory with
 the local nature has been ``improved'' by the envelope equation
 to become a global solution.

There is another version of the RG method\cite{goldenfeld}, which
 involves a ``renormalization'' of the parameters.
We shift the parameter for the local curves as follows:
Let $\tau$ be close to $t$,  and
write
 $t - t_0 = t-\tau +\tau - t_0$. Then putting that
\beq
A(\tau)&=& A_0(t_0)Z(t_0, \tau),\ \
Z(t_0, \tau)= 1 - \frac{\eps}{2}(\tau -t_0) + \frac{\eps^2}{8}
 (\tau -t_0)^2,\nonumber \\
 \theta(\tau)&=&\theta_0(t_0)+ \delta\theta, \ \ \
 \delta\theta = -  \frac{\eps^2}{8}(\tau -t_0),
\eeq
we have
\beq
x(t, \tau)&=& A(\tau)\sin (t +\theta(\tau)) -\eps\frac{A(\tau)}{2}
 (t -\tau)\sin(t+\theta(\tau))
  \nonumber \\
 \ \ \ & \ \ \ & + \eps^2\frac{A(\tau)}{8}
\{ (t-\tau)^2\sin(t +\theta(\tau)) - (t-\tau)\cos(t+\theta(\tau))\}
  + O(\eps^3),
\eeq
where
\beq
x(\tau, \tau)= A(\tau)\sin (\tau +\theta(\tau)).
\eeq
Then the envelope of the curves given by $\{x(t, \tau)\}_{\tau}$
 will be found to be the same as given in Eq. (3.12).

This may concludes the account of our formulation of the RG method
based on the
classical theory of envelopes.
 However,  there is a problem left:
Does $x_{_E}(t)\equiv x(t,t)$ indeed satisfy the original differential
equation?
 In our simple example,  the result Eq.(3.12) shows
 that it does.
 It is also the case for  all the  resultant solutions
worked out here and in
\cite{goldenfeld}.
 We are, however, not aware of a general proof available to
show that the envelope function should
 satisfy the differential equation (uniformly)
up to the same order as the local
 solutions do locally.
 We give here a proof for that for a wide class of linear and
 non-linear ordinal differential
equations (ODE). The proof can be easily generalized to  partial differential
equations (PDE).

\newcommand{\bfq}{{\bf q}}
{\newcommand{\bff}{{\bf F}}

Let us assume that the differential equation under consideration can be
 converted to the following coupled equation of {\em first order}:
\beq
\frac{d{\bf q}(t)}{dt}={\bf F}({\bf q}(t), t; \eps),
\eeq
where \ $^t\bfq =(q_1, q_2, \cdots )$  and $\bff$  are column vectors.
 It should be noted  that  ${\bf F}$ may be a
 non-linear function of $\bfq$ and $t$, although
in our example,
\beq
q_1=x,\ \ \  q_2 =\dot{x}, \ \ \ {\bf F}=
\pmatrix{ q_2 \cr
       -q_1 -\eps q_2},
\eeq
i.e., $\bff$ is linear in $\bfq$.
We also assume that we  have an approximate local solution
 $\tilde{\bfq}(t, \ t_0)$
 around $t=t_0$ up to $O(\eps^n)$;
\beq
\frac{d\tilde{{\bf q}}}{dt}={\bf F}(\tilde{{\bf q}}, t; \eps)\ + O(\eps^n).
\eeq
One can see for our example to satisfy  this using Eq.(3.6).

The envelope equation implies
\beq
\frac{\partial \tilde{\bfq}(t, t_0)}{\partial t_0}=0
\eeq
at $t_0=t$.  With this condition,
  $\bfq_{_E}(t)$ corresponding to $x_{_E}(t)$ is defined by
\beq
\bfq_{_E}(t)=\tilde{\bfq}(t, \ t).
\eeq

It is now easy to show that $\bfq_{_E}(t)$ satisfies Eq.(3.16) up to the
 same order as $\tilde{\bfq}(t, \ t_0)$ does:
 In fact, for  $ \forall t_0$
\beq
\frac{d\bfq _{_E}(t)}{d t}\Biggl\vert_{t= t_0}=
\frac{d\tilde{\bfq}(t, t_0)}{dt}\Biggl\vert _{t=t_0}
 + \frac{\partial \tilde{\bfq}(t, t_0)}{\partial t_0}\Biggl\vert_{t=t_0}
 = \frac{d\tilde{\bfq}(t, t_0)}{dt}\Biggl\vert _{t=t_0},
\eeq
where Eq.(3.19) has been used.
And noting that ${\bf F}({\bf q}_{_E}(t_0), t_0; \eps)=
{\bf F}({\bf q}(t_0,t_0), t_0; \eps)$, we  see for $\forall t$
\beq
\frac{d{\bf q}_{_E}}{dt}={\bf F}({\bf q}_{_E}(t), t; \eps)\ + O(\eps^n),
\eeq
 on account of Eq. (3.18). This completes the proof.
  It should be stressed that Eq.(3.22) is valid
 uniformly for $\forall t$ in contrast to Eq.(3.18) which is valid only
 locally around $t=t_0$.

\setcounter{equation}{0}
\section{Examples}
\renewcommand{\theequation}{\thesection.\arabic{equation}}

Let us take a couple of examples to apply our formulation.
 These can be converted to equations in  the form given in Eq. (3.16).

\subsection{A boundary-layer problem}

The first example is a typical   boundary-layer problem\cite{bender}:
\beq
\eps \frac{d^2y}{dx^2} + (1+ \eps) \frac{dy}{dx} + y=0,\ \
\eeq
with the boundary condition $y(0)=0, \ \ y(1)=1$. The exact solution
 to this problem is readily found to be
\beq
y(x)=\frac{\exp(-x) - \exp(-x/\eps)}{\exp(-1) - \exp(-1/\eps)}.
\eeq

Now let us solve the problem in the perturbation theory.
Introducing the inner variable $X$ by $\eps X=x$ \cite{bender}, and putting
 $Y(X)=y(x)$,  the equation is
converted to the following;
\beq
 \frac{d^2Y}{dX^2} + \frac{dY}{dX} =\ -\eps (\frac{dY}{dX}+ Y).
\eeq
Expanding $Y$ in the power series of $\eps$ as
$Y = Y_0 + \eps Y_1+\eps^2 Y_2 + ... $, one has
\beq
Y_0'' + Y_0'&=&0,\nonumber \\
Y_1'' + Y_1'&=&- Y_0'- Y_0,\nonumber \\
 \ \ \ & \vdots . &
\eeq
Here, \ $Y'\equiv dY/dX$ etc. To solve the equation, we set a boundary
 condition to $Y(X)$ and $Y_0(X)$ at $X=X_0$;
\beq
 Y(X)=Y_0(X_0)= A_0,
\eeq
 where  $X_0$   is  an arbitrary constant  and $A_0$ is supposed to be a
function of $X_0$.

For this problem, we shall follow the prescription given in
\cite{goldenfeld} for the higher order terms.
 Then the solutions to these equations may be written as
\beq
Y_0(X)&=& A_0 - B_0\e^{ -(X-X_0)},\nonumber \\
Y_1(X)&=& -A_0(X-X_0)- (B_0+C_0)(\e^{-(X-X_0)} -1).
\eeq
Defining $A=A_0 + \eps(B_0+C_0)$ and $B=B_0 + \eps (B_0+C_0)$, we have
\beq
Y(X, X_0)= A -B\e^{-(X-X_0)} -\eps A(X-X_0) + O(\eps^2).
\eeq
In terms of the original coordinate,
\beq
y(x, x_0)= Y(X, X_0)= A -B\e^{-(x-x_0)/\eps} - A(x-x_0) + O(\eps^2),
\eeq
 with $x_0 =X_0/\eps$.

Now let us obtain the envelope $Y_{_E}(X)$
 of the family of functions $\{Y(X, X_0)\}_{X_0}$
 each of  which has the common tangent with  $Y_{_E}(X)$ at $X=X_0$.
 According to the standard procedure to obtain the envelope,
 we first solve the equation,
\beq
\frac{\partial Y}{\partial X_0}=0,\ \ \ {\rm with}\ \ X_0=X,
\eeq
and then identify as $Y(X, X)= Y_{_E}(X)$.

Eq. (4.9) claims
\beq
A' + \eps A=0, \ \ \ B' + \eps B=0,
\eeq
with the solutions $A(X)= \bar{A}\exp(-\eps X),\ \ B(X)=\bar{B}\exp(-X)$,
where $\bar{A}$ and $\bar{B}$ are constant.
Thus one finds
\beq
Y_{_E}(X)= Y(X,X)= A(X) - B(X)= \bar{A}\e^{-\eps X} - \bar{B}\e^{-X}.
\eeq
In terms of the original variable $x$,
\beq
y_{_E}(x) \equiv Y_{_E}(X)= \bar{A}\exp(-x) - \bar{B}\exp(-\frac{x}{\eps}).
\eeq
It is remarkable that the resultant $y_{_E}(x)$ can admit both the inner
  and
 outer boundary conditions simultaneously; $y(0)=1, \ \ y(1)=1$.
 In fact, with the
boundary conditions we have $\bar{A}=\bar{B}=1/(\exp(-1) - \exp(-1/\eps))$,
 hence $y_{_E}(x)$ coincides with the exact solution $y(x)$ given in
 Eq. (4.2).

In Fig. 2, we show the exact solution $y(x)$
and the local solutions $y(x, x_0)$ for several $x_0$:
 One can clearly see that the exact solution is the envelope of
 the curves given by $\{y(x, x_0)\}_{x_0}$.

\vspace{1cm}
{\cl {\fbox{\bf Fig.2}}}

A comment is in order here:  If we adopted  the prescription given in \S 3
 for the higher order terms,
the perturbed solution $Y_1(X)$ reads $Y_1(X)= -A_0(X-X_0)$; note
 the boundary condition Eq.(4.5). Then the
proceeding calculations after Eq. (4.6) would be slightly simplified.

\subsection{A non-linear oscillator}

In this subsection, we consider the following Rayleigh
equation\cite{bender,goldenfeld},
\beq
\ddot{y} + y = \eps ( \dot{y} - \frac{1}{3}\dot{y}^3).
\eeq

 Applying the perturbation theory with the expansion
 $y= y_0 + \eps y_1 + \eps^2 y_2 + \cdots$,
one has
\beq
y(t, t_0)&=& R_0 \sin(t +\theta_0)+ \eps\{
  (\frac{R_0}{2} - \frac{R_0^3}{8})(t - t_0) \sin(t + \theta_0)\nonumber \\
 \ \ & \ & \hspace{2cm}    +
\frac{R_0^3}{96}(\cos3(t+\theta_0)  \}+ O(\eps^2).
\eeq
Here we have not included the terms proportional to the
unperturbed solution in the higher order terms
 in accordance with the prescription given in \S 3, so that the following
 calculation is somewhat simplified than in \cite{goldenfeld}. Furthermore,
 the result with this prescription will coincide with the one given in the
 Krylov-Bogoliubov-Mitropolsky method\cite{kbm}, as we will see in
 Eq. (4.18).

Eq. (4.14) gives a family of curves $\{C_{t_0}\}_{t_0}$
parametrized with $t_0$.
The envelope $E$ of $\{C_{t_0}\}_{t_0}$ with the point of tangency at
 $t=t_0$ can be obtained as follows:
\beq
\frac{\partial y(t, t_0)}{\partial t_0}=0,
\eeq
with $t_0=t$.  Assuming that $\dot{R_0}$ and $\dot{\theta}_0$
 are  $\sim O(\eps)$ at most,
 we have
\beq
\dot{R}_0= \eps (\frac{R_0}{2} - \frac{R_0^3}{8}),
 \ \ \ \ \dot{\theta}_0=0,
\eeq
the solution of which reads
\beq
R_0(t)=\frac{\bar{R}_0}
 {\sqrt{\exp(- \eps t) + \bar{R}_0^2( 1- \exp( -\eps t))/4}},
\eeq
 with $\bar{R}_0=R_0(0)$ and $\theta_0=$ \ constant.
Thus the envelope is given by
\beq
y_E(t)=y(t, t)= R_0(t)\sin(t +\theta_0)+
\eps \frac{R_0(t)^3}{96}(\cos3(t+\theta_0))+ O(\eps^2).
\eeq
This is  an approximate but global solution to Eq. (4.13) with a limit
 cycle in accordance with the result given in \cite{kbm}.  We note that
 since Eq.(4.13) can be rewritten in the form of Eq.(3.16), Eq.(4.18)
 satisfies Eq.(4.13) up to $O(\eps^2)$.

\setcounter{equation}{0}
\section{A brief summary and concluding remarks}
\renewcommand{\theequation}{\thesection.\arabic{equation}}

We have given a geometrical formulation of  the RG method
 for global analysis
 recently proposed by Goldenfeld et al\cite{goldenfeld}:
 We have shown that the RG equation can be interpreted as an envelope
 equation, and  given a purely
mathematical foundation to the method. We have also given a proof that
the envelope function satisfies the same differential
 equation up to the same order
 as the functions representing the local curves do.

  It is important  that
 a geometrical meaning of the RG equation even in a generic sense
 has been clarified in the present work.
The RG equation appears in various
 fields in physics. For example, let us take a model in the quantum field
 theory\cite{coleman};
\beq
{\cal L}= \half (\partial_{\mu}\phi)^2 - \frac{\lambda}{4!}\phi^4 +
{\rm c.t.},
\eeq
where c.t. stands for counter terms.
 The true vacuum in the quantum field theory is determined by the minimum
 of so called the effective potential
${\cal V}(\phi_c)$\cite{coleman,jona}. In the
 one-loop approximation, the renormalized effective potential reads
\beq
{\cal V}(\phi_c, M)= \frac{\lambda}{4!}\phi_c^4 +
\frac{\lambda^2\phi_c^4}{256\pi ^2}
(\ln \frac{\phi_c^2}{M^2} - \frac{25}{6}),
\eeq
where $M^2$ is  the renormalization point.
 To  see a correspondence to the envelope theory,
 one may parametrize as $\phi_c^2={\rm exp}\ t$ and $M^2={\rm exp}\ t_0$ ,
 then  one sees that
$\ln {\phi_c^2}/{M^2}$ becomes  a secular term $t- t_0$.
In the quantum field theory, one applies the RG equation to
 improve the effective potential as follows\cite{coleman,kugo};
\beq
\frac{\partial {\cal V}}{\partial M^2}=0, \ \ \ {\rm with } \ \
M^2 = \phi_c^2.
\eeq
One sees that this is the envelope equation! The resultant ``improved''
 effective potential is found to be
\beq
{\cal V}_{\rm impr}(\phi_c)= {\cal V}(\phi_c, \phi_c)=
 \frac{\frac{\lambda}{4!}\phi_c^4}{1 - \frac{3\lambda}{16\pi^2}
 \ln\frac{\phi_c}{\phi_{c0}}}.
\eeq
Thus one can now understand  that the ``improved'' effective
 potential is
 nothing but   the envelope of the effective potential in the
 perturbation theory. One also sees the reason why the RG equation with
 $\phi_c=M$ can ``improve'' the effective potential.
 Then what is the physical significance of the envelope function
  ${\cal V}_{\rm impr}(\phi_c)$?
 One can readily show that for $\forall M$
\beq
\frac{\partial {\cal V}(\phi_c)_{\rm impr}}{\partial \phi_c^2}\Biggl\vert
_{\phi_c =M}=
\frac{\partial {\cal V}(\phi_c, M)}{\partial \phi_c^2}\Biggl\vert_{\phi_c=M},
\eeq
 owing to the envelope condition Eq. (5.3).  This implies, for example,
that the  vacuum condensate $\phi_c$ that is given by
$\partial {\cal V}_{\rm impr}/{\partial \phi_c^2}=0$ is
 correct up  to the same order of $\hbar$-expansion in which the original
 effective potential is calculated; this is
 irrespective of how large is the resultant $\phi_c$.
  Detailed discussions of the
application of the envelope theory to the quantum field theory
will be reported elsewhere\cite{effpot}.\footnote{Recent renewed interest in
 ``improving'' the effective potential is motivated by  the problem of
 how to ``improve'' the effective potentials with multi scales as appear
 in the standard model\cite{kugo}.
 The observation given here that  the RG equation can be
 interpreted  as the
envelope equation may give an insight into  how to construct
 effective potentials with a global nature in multi-scale cases.}

The RG equation has also a remarkable success
 in statistical physics especially in the critical phenomena \cite{JZ}.
 One may also note that there is another successful theory of the critical
 phenomena called coherent anomaly method (CAM)\cite{CAM}.
 The relation between CAM and the RG equation theory is not known.
  Interestingly enough, CAM utilizes
 {\em envelopes} of susceptibilities and other thermodynamical
 quantities as a function of temperature.
 It might be possible to give a definite relation between CAM and
the RG theory because the RG equation can be interpreted as an
 envelope equation,  as shown in this work.

Mathematically, it is most important to give a rigorous proof for the
 RG method in general situations and to clarify what types of
differential equations can be analyzed in this method, although we have given
 a simple proof for a class of ODE's. We note that
 the proof can be generalized to  partial differential
 equations, especially of first order with respect to a
variable\cite{pde}.
One should be also able to estimate the accuracy  of the envelope theory
 for a given equation.  We hope that this paper may stimulate
 studies for a deeper understanding of global analysis based on the theory of
 envelopes.

\vspace{2cm}
{\cl {\large{\bf Acknowledgements}}}

The author got to know the RG method by the lectures given
by G. C. Paquette
 at  Ryukoku University in June 1994. The author acknowledges him for
 his pedagogical lectures.
 The author thanks Y. Morita, who as a mathematician
 did not pretend to understand the author's explanation of the RG method.
This incident  prompted the author
 to think about  the mathematical meaning of the  RG equation.
The author is grateful to M. Yamaguti and T. Kugo for their
 interest in this work.

\newcommand{\NG}{N. \ Goldenfeld}
\newcommand{\YO}{Y.\ Oono}

\newpage
{\large {\bf Figure Captions}}
\begin{description}
\item{Fig.1}  A family of functions and its envelope:\\
 The thin lines show
$y=\exp(-\eps \tau_0)( 1- \eps(x -\tau_0)) + \exp(-x)$
 with $\tau_0 = 0.2,\ 0.4,\ 0.4, 0.6$ and $0.8$, which are attached to
 the respective lines.  The thick line shows
 the envelope $y=\exp(-\eps x) -\exp(-x)$.\ \ ($\eps = 0.8$.) \\

\item{Fig. 2}
The thin lines show $y(x, x_0)= A(x_0)- B(x_0)\exp(-(x-x_0)/\eps)
- A(x_0)(x-x_0)$ with $x_0=0.2, 0.4 $ and $0.8$, which are attached to the
 respective lines.  The thick line
 shows $y(x)$ given in Eq.(4.2). ($\eps = 0.1$.)
\end{description}

\end{document}